\documentclass[12pt]{iopart}

\usepackage{comment}
\usepackage{graphicx}
\usepackage{hyperref}
\usepackage{color}

\begin{document}

\title[]{Investigating the Kinetic Effects on Current Gradient-Driven Instabilities of Electron Current Layers via Particle-in-Cell Simulations}

\author{Sushmita Mishra$^1$, \footnote{Corresponding author (email: gurudatt.gaur@sxca.edu.in)}Gurudatt Gaur$^2$, Bhavesh G. Patel$^3$}

\address{$^1$Institute of Advanced Research, Gandhinagar 382 426, India} 
\address{$^2$St. Xavier's College (Autonomous), Navrangpura, Ahmedabad 380 009, India}
\address{$^3$Institute for Plasma Research, Bhat, Gandhinagar 382 428, India}
%\ead{mishra.sushmita0@gmail.com}
%\ead{gurudatt.gaur@sxca.edu.in}
%\ead{bhavesh@ipr.res.in}

\vspace{10pt}
\begin{indented}
\item[\today] 
\end{indented}

\begin{abstract}
Electron current layers, which form in various natural and laboratory plasmas, are susceptible to multiple instabilities, with tearing being a prominent instability driven by current gradients. Tearing is considered a potential mechanism for magnetic reconnection in collisionless regimes, where electron inertia acts as a non-ideal factor that causes magnetic field lines to break and reconnect. In contrast, another mode, known as the surface-preserving mode, also driven by current gradients, maintains the magnetic field topology. In this study, we investigate the kinetic effects on these modes in the presence of finite electron temperatures using two-dimensional particle-in-cell simulations. Our findings reveal that temperature significantly stabilizes the tearing mode, particularly at higher temperatures, due to an increased electron Larmor radius and the associated magnetic field diffusion. We also examine the interplay between the guide field and temperature. Additionally, we observe that growth rates for the surface-preserving mode, in contrast to the tearing mode, increase with temperature, likely due to enhanced electron flow velocities. Furthermore, we identify cases with mixed modes, where both tearing and surface-preserving modes coexist, exhibiting asymmetric structures characteristic of asymmetric magnetic reconnection. Finally, we outline potential future research directions that build upon our findings.

\end{abstract}

%\maketitle

\section{Introduction}

Plasma is a state of matter in which electrons and ions move, exhibiting collective behavior under the influence of electric and magnetic fields generated either by their own motion or by external sources. Magnetic fields significantly influence plasma dynamics, leading to rich and complex behaviors. They affect the motion of charged particles, which, in turn, modify the magnetic fields due to the ``frozen-in" condition arising from plasma’s high electrical conductivity. A key phenomenon where the magnetic field is altered by plasma particles is magnetic reconnection (MR). In MR, plasma pushes two oppositely directed magnetic field lines towards each other, changing the topology of the magnetic field lines \cite{birnbook, biskamp}. MR occurs in various natural settings, including within the solar system \cite{yamada}. It plays a crucial role in solar flares, coronal mass ejections, the Earth's magnetosphere, and astrophysical jets, and is also believed to occur during star formation. Additionally, MR is observed in several fusion plasma devices, such as Tokamak \cite{mai}, Reverse Field Pinch \cite{porcu}, and Spheromak configurations \cite{taylor1986, yamada1999b}. Numerous laboratory plasma devices have been developed to investigate the mechanisms behind MR \cite{stenzel1, stenzel2, sayak}.

Significant efforts have been made to propose theoretical models that explain MR at different time and length scales in various plasma conditions. The resistive MHD (magnetohydrodynamics) models, such as the Sweet-Parker and Petschek models{\cite{ji}, describe MR occurring in long, narrow current sheets with a reconnection rate limited by plasma resistivity. The Sweet-Parker model predicts a slow reconnection rate, insufficient to explain the fast reconnection events observed in nature. In contrast, the Petschek model suggests a faster reconnection process, although it is less applicable to collisionless plasmas\cite{yamada}. In space and astrophysics contexts, where fast reconnection occurs, plasmas are known to be collisionless. In a collisionless plasma, where the mean free path of particles is much larger than the system size, reconnection dynamics are predominantly governed by electromagnetic forces rather than collisional interactions. 

Various models have been proposed to understand fast reconnection in collisionless plasmas, addressing different aspects and scales of the process, both fluid and kinetic. While the MHD model is valuable for understanding large-scale plasma behavior, it is insufficient for describing the detailed physics of collisionless MR. For collisionless plasmas, models that include electron-scale dynamics, such as EMHD (electron magnetohydrodynamics) \cite{bulanov1992, gaur2016}, Hall MHD \cite{birnbook,yamada}, and two-fluid models \cite{ shay, drake2008, biskamp}, as well as models that incorporate kinetic effects, such as hybrid \cite{matthews} and full kinetic (PIC: particle-in-cell) models \cite{daughton, horiuchi}, are necessary to accurately capture the complex dynamics and fast reconnection rates observed in these environments. These models incorporate the Hall term and electron inertia, leading to faster reconnection rates and more complex magnetic island structures than single-fluid MHD models can provide.

A specific manifestation of magnetic reconnection (MR) is the tearing instability, a dynamic approach to studying MR. Unlike steady-state approaches exemplified by the Sweet-Parker and Petschek models, the tearing instability involves the formation of a chain of magnetic islands and reconnection layers, leading to a reconfiguration of the magnetic topology. The tearing mode evolves according to parameters such as the width of the current sheet ($\epsilon$) and length scales characteristic of kinetic effects, including the electron inertial skin depth ($d_e$), ion inertial skin depth ($d_i$), and Larmor radius of electrons or ions ($r_L$) \cite{biskamp}. The tearing instability in thin current layers with thickness smaller than the ion skin depth has been widely studied in the context of 2D and 3D EMHD.

Gaur and Kaw \cite{gaur2016} investigated the susceptibility of thin electron current sheets to tearing and surface-preserving instabilities within 2D EMHD when equilibrium length scales are shorter than the ion skin depth. These instabilities, driven by current gradients\footnote{For a discussion on the role of current gradient, the reader is referred to the Appendix.}, exhibit distinct behaviors: the tearing mode results in magnetic island formation, while the surface-preserving mode maintains magnetic flux surfaces. The tearing mode is non-local, unstable to larger-scale perturbations, while the surface-preserving mode is local, unstable to shorter wavelength perturbations. Similar studies by Lukin \cite{lukin2009} and Jain et al. \cite{jain2015, jain2017} have examined these modes within the EMHD framework.

Building on Gaur and Kaw's work, we explore the kinetic effects on tearing and surface-preserving modes, specifically the influence of temperature and the interplay between temperature and guide fields, using 2D PIC simulations. To our knowledge, this is the first study to address these modes within a kinetic framework.

Fluid descriptions such as EMHD are inadequate at kinetic scales, particularly for high-temperature plasmas or when length scales approach the particle gyroradius. In situations dominated by high-frequency phenomena, strong magnetic fields, turbulence, or collisionless dynamics, a kinetic approach is necessary. PIC simulations, which treat particles individually rather than collectively as a fluid, allow us to capture the intricate kinetic behavior of plasmas \cite{dawson1983}. In this study, we utilize PIC simulations to examine the linear and nonlinear evolution of tearing and surface-preserving instabilities in response to kinetic effects.

This paper is organized as follows. In Section II, we discuss the basics of the particle-in-cell scheme along with the simulation setup used to model tearing and surface-preserving instabilities in electron current layers (representing a thin electron current sheet). In Section III, we present and analyze the results from various simulation runs, focusing on the behavior of the tearing and surface-preserving modes in the presence of temperature. This section also includes a comparison with existing studies to highlight the novelty of our work. In Section IV, we provide conclusions and discuss future prospects of our research.

\section{Simulation Setup}

Various PIC codes \cite{dieckmann2024} are available for plasma simulations, with OSIRIS being one of the most widely used. OSIRIS is a state-of-the-art PIC simulation code that is fully relativistic, massively parallel and fully object-oriented. Written in Fortran-90, it is designed for high-performance computing with multi-dimensional capabilities \cite{fonseca}. 

We have utilized the OSIRIS 4.0 framework \cite{fonseca2018osiris,fonseca2013,kemp,silva} to model the 2D tearing and surface preserving instabilities. In our simulations, the normalized units are defined as follows: length is normalized by electron skin depth c/$\omega_{pe}$, time is normalized by $\omega_{pe}^{-1}$, velocity is normalised by c and magnetic field by the quantity $\frac{m_e c^2}{e(c/\omega_{pe})}$. Where c is the speed of light, $\omega_{pe}$ is the electron plasma frequency, and $m_e$ is the electron mass. In this study, all physical quantities are expressed in cgs units.

In our simulations, we use a Cartesian coordinate system where the axes are defined as follows:  the x-axis represents the shear or gradient direction, corresponding to the current sheet gradient; the y-axis aligns with the equilibrium magnetic field (Beq), representing a uniform in-plane magnetic field; and the z-axis corresponds to the symmetry or out-of-plane direction. This z-axis is associated with equilibrium velocity/current density, and also serves as the direction of the applied guide magnetic field \(B_G\). The coordinate system and field orientations are depicted schematically in Fig.~\ref{fig:diagram}.

\begin{figure}[h!]
\centering
\includegraphics[scale=0.3]{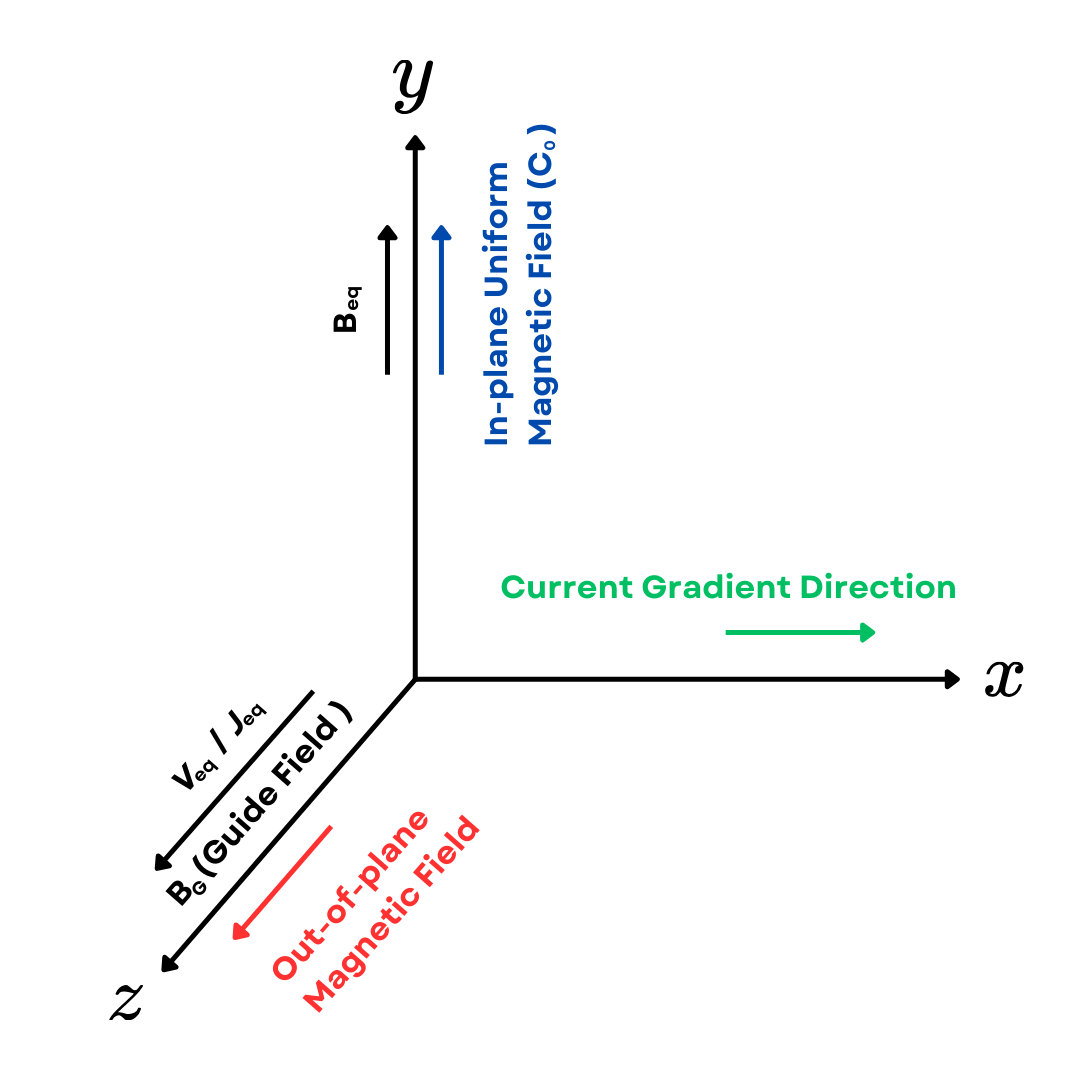}% Here is how to import EPS art
\caption{\label{fig:diagram} Schematic representation of the coordinate system and field orientations in the simulation setup.}    
\end{figure}

We have initialized our simulations for a thin current sheet equilibrium having width of the magnitude $\epsilon$
and $z$ is the symmetry direction i.e., $(\frac{\partial}{\partial z})$ = 0. Ions are at rest ({\it i.e.}, $v_i$ = 0). Equilibrium magnetic field is expressed as follows:

\begin{eqnarray}
\textbf{B}_{eq} ~=~ \hat{y}B_0(x) ~=~ \hat{y} (B_{00}\{sech(x/\epsilon) tanh(x/\epsilon)\} ~+~ C_0),
\end{eqnarray}
Corresponding equilibrium velocity is written as,
\begin{eqnarray}
\textbf{v}_{eq} \left(=-\frac{\nabla \times \textbf{B}_{eq}}{n_0e} \right)  ~&=&~ \hat{z}v_0(x)  \nonumber \\
~&=&~ \hat{z}\frac{B_{00}~ sinh^2 (x/\epsilon)}{(n_0 e) \, \epsilon \, cosh^3(x/\epsilon)}    
\end{eqnarray}
Where, $\epsilon$ is shear width, $B_{00}$ is the amplitude of the magnetic field, $C_0$ is a uniform magnetic field added externally in the direction of field.

\begin{figure}
\centering
\includegraphics[scale=0.55]{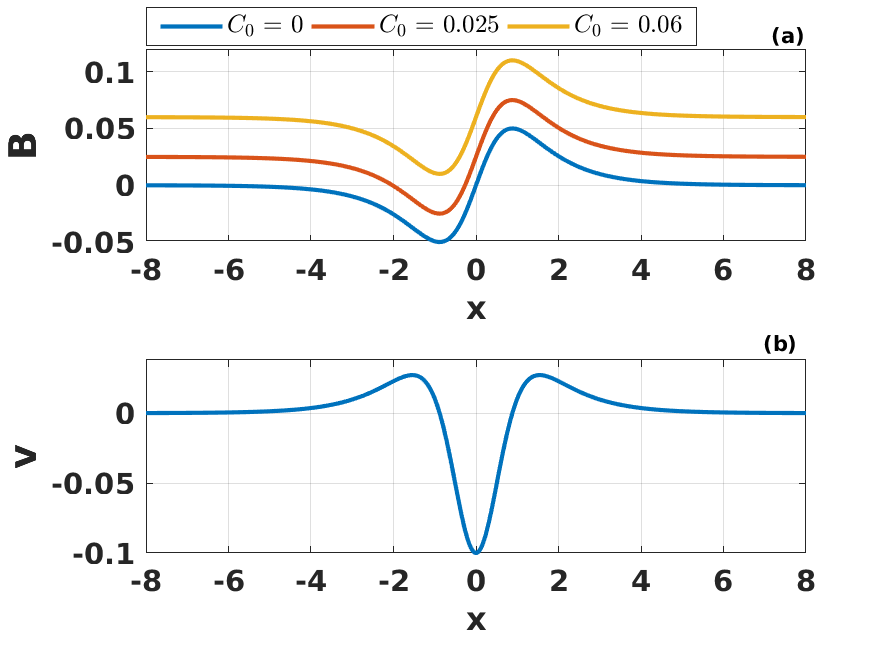}
\caption{\label{fig:profiles} (a) Equilibrium magnetic field and (b) equilibrium velocity profiles.}
\end{figure}

In all our simulation runs, we have chosen $B_{00} = 0.1$ and $\epsilon = 1$.
%The equilibrium magnetic field has been chosen in such a way that the quantity $B_{0}-B_{0}^{''}$ changes sign which is necessary for any mode of tearing instabilities to exist. 
A 2D rectangular simulation box with dimensions $L_x = 8(c/\omega_{pe})$ and $L_y = 3(c/\omega_{pe})$ has been chosen. The spatial resolution corresponds to a grid size of $\Delta x = 2L_x / 512$ and $\Delta y = 2L_y / 256$. The total number of particles per cell is 50. We allow the equilibrium to evolve against the numerical perturbations. We have taken a periodic boundary conditions for both fields and particles because the equilibrium profile chosen is such that it ceases at both the boundaries. The equilibrium magnetic field and the corresponding equilibrium velocity profiles are shown in Fig.~\ref{fig:profiles}. Three different magnetic field profiles correspond to three different values of $C_0$ {\it viz.}, 0, 0.025, 0.06.

\section{Results and Discussion}

In this section we present and discuss the results from various simulation runs obtained by varying the $C_0$ (a uniform magnetic field applied along the equilibrium magnetic field) and $T$ (temperature).%\\ \\

\subsection{$C_0 = 0$: Only Tearing Mode Case }
We begin by setting \( T = 0 \), corresponding to the typical tearing instability case, which has been extensively studied by various authors using both fluid and PIC simulations. We reproduced this scenario to validate our simulation setup. In Fig.~\ref{fig:pte}, we plot the evolution of the perturbation energy over time. During the initial phase of the simulation, the perturbation energy grows exponentially. The slope of this portion (a straight line on the semilog plot) closely matches twice the maximum growth rate \( 2\gamma \) derived from linear instability calculations \cite{gaur2016}. The dashed line alongside the simulation curve has a slope equal to \( 2\gamma \), showing good agreement. As the amplitude of the perturbed field increases, nonlinear effects become significant, leading to saturation of the perturbation energy in the later stages of the simulation.

In Fig.~\ref{fig:streamlines_0eV}, the contours in the background depict the structure of the out-of-plane magnetic field. The magnetic field lines, superimposed on these contours, are straight, initially. However, as the tearing instability progresses, these field lines begin to bend, eventually forming an island structure. Concurrently, the out-of-plane magnetic field develops a distinctive quadrupolar pattern. These features are characteristic indicators of the tearing instability.
\begin{figure}
\centering
\includegraphics[scale=0.55]{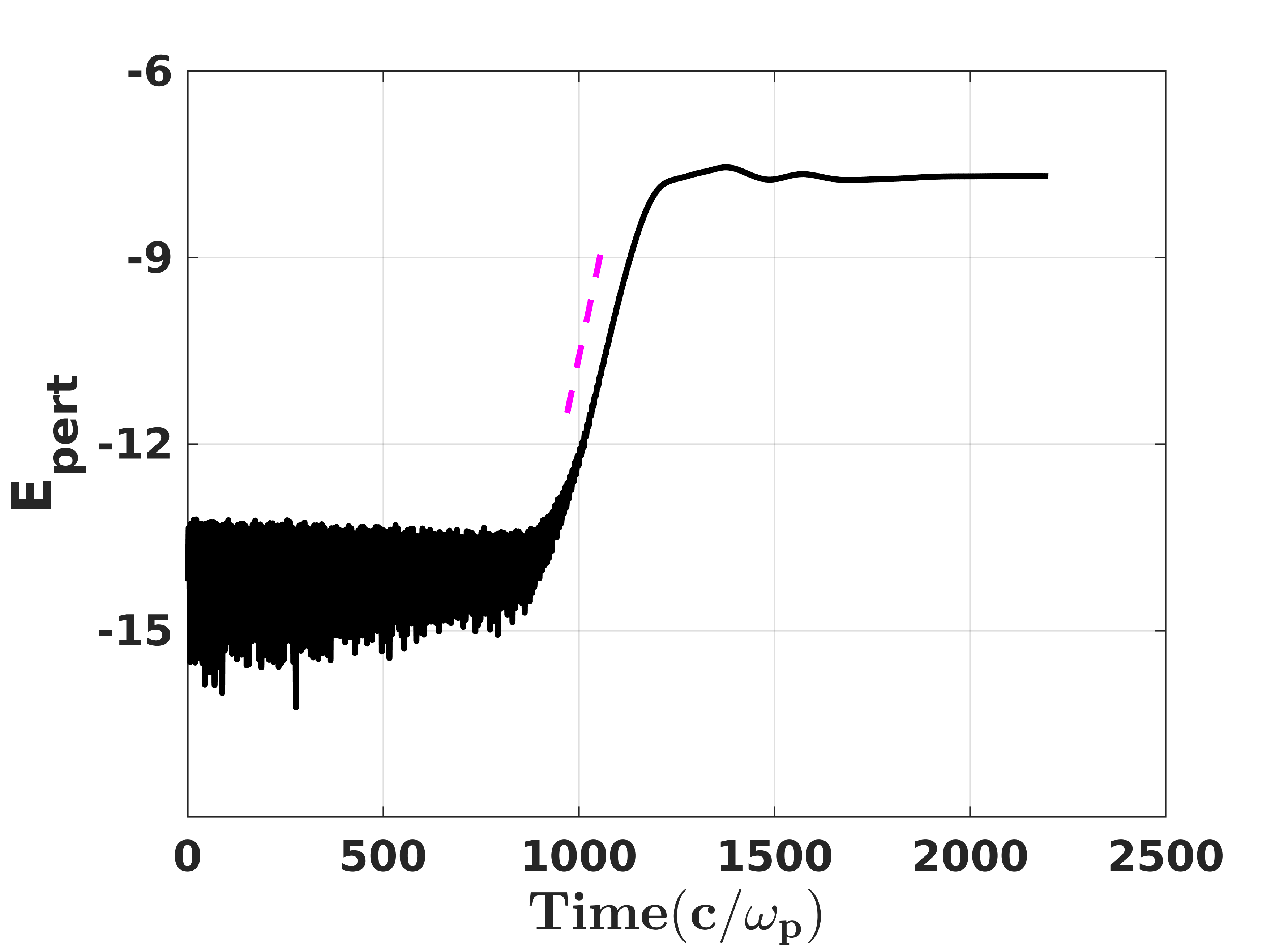}% Here is how to import EPS art
\caption{\label{fig:pte} Evolution of perturbed energy over time. The dashed straight lines drawn alongside has a slope of $2\gamma$, where $\gamma = 0.0188$ is the linear growth rate obtained from the linear analysis.}
\end{figure}

\begin{figure}
\centering
\includegraphics[scale=0.90]{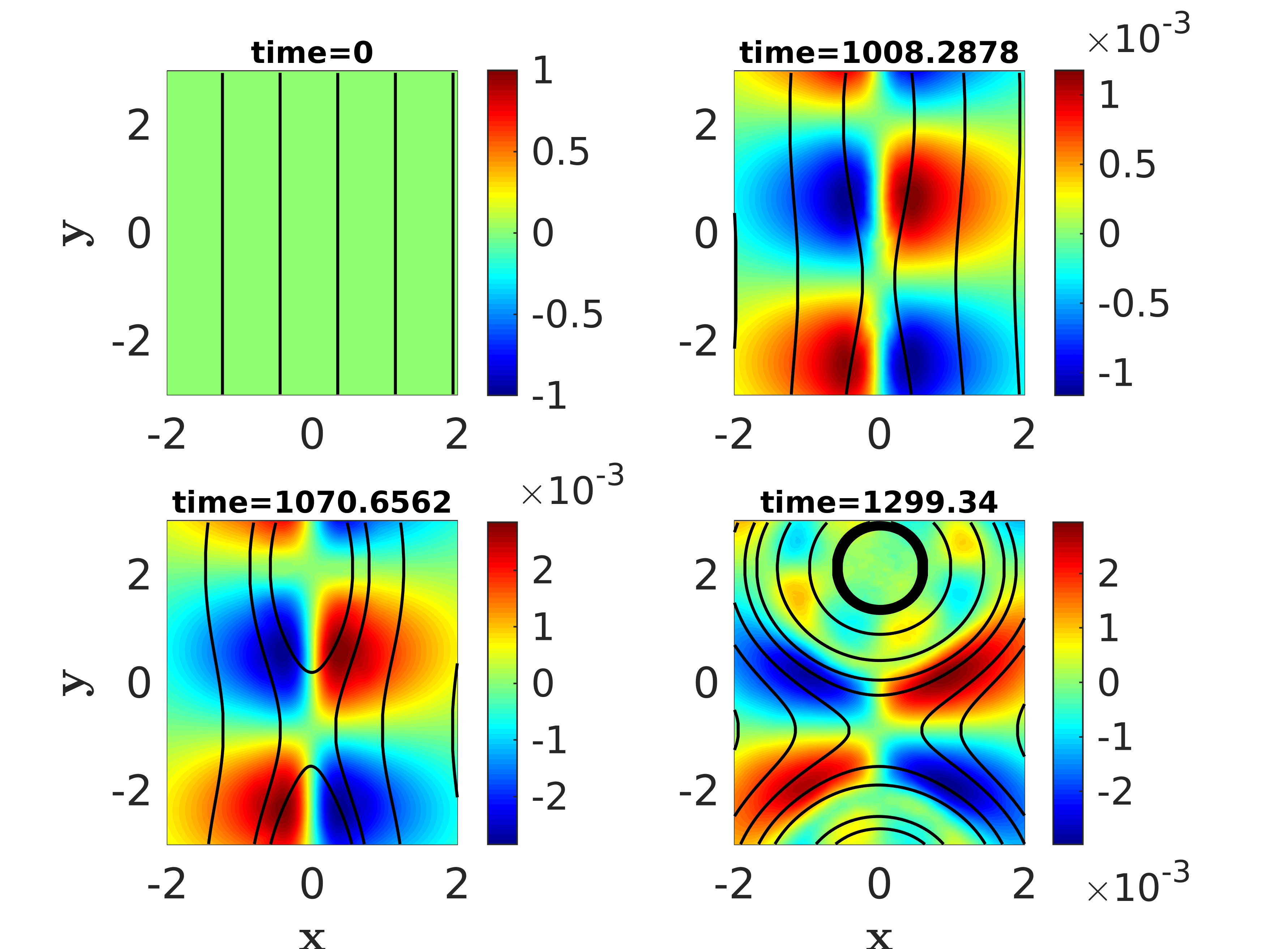}% Here is how to import EPS art
\caption{\label{fig:streamlines_0eV} Shaded isocontours of out-of-plane magnetic field (\( B_z \)) with superimposed streamlines (solid lines) over various times for pure tearing mode case ($C_{0}=0$) and T = 0 eV case. The color bar indicates the normalized amplitude of \(B_z\).}
\end{figure}

Having validated the simulation setup, we now proceed to study the effect of temperature on tearing instability. We conduct simulation runs for the following temperature values: $( T = 5, 10, 20, 50, 100, 200, 500, 1000)$ electron volts. Upon increasing the temperature, while the characteristic features of the instability (such as the formation of islands and quadrupoles) remain more or less the same, we observe a variation in the growth rate with temperature. Fig.~\ref{fig:pt_GR vs T} illustrates the variation in the growth rate as temperature increases. The growth rate initially increases with temperature and then starts decreasing monotonically. This behavior can be understood as follows.

Initially, when the temperature increases (say, up to 10 eV), the kinetic energy of the electrons increases. This makes the tearing instability more favorable, as the electrons move faster in the tearing instability plane compared to the case with \( T = 0 \). Consequently, the instability growth rate increases marginally. If the temperature further increases (beyond 10 eV), electron larmor radius ($r_L$ ) increases significantly, which is given as $r_L$ = $v_{th}/\omega_{ce}$, where $v_{th}$ = $\sqrt{p_e/(nm)}$ and $\omega_{ce} = eB/(m_{e}c)$ are the thermal velocity and plasma cyclotron frequency respectively. In other words, kinetic pressure starts dominating the the magnetic pressure i.e., the plasma beta ($\beta$) value increases, where $\beta=(n_ek_BT/(B^2/2\mu_0))$. This increases the diffusion of magnetic field (not just near the null-line) which, in turn, reduces the strength of current-gradient (due to weaker magnetic field and hence weaker current). As a result, the growth rate decreases. Eventually, at T $\ge 1 keV$, the instability completely suppresses.

To reinforce our argument on the role of larmor radius in tearing growth rate against temperature, we add a uniform guide field \footnote{Presence of a guide field is known to play a stabilizing role in magnetic reconnection.} (in the direction perpendicular to the plane of tearing mode). Addition of guide field increases the magnetic pressure (energy) of the system. Owing to which, the decrease in the growth rate now start dropping at a higher temperature. Fig.~\ref{fig:guide field} shows that as the guide field increases, the peak in the growth rate shift towards higher temperature. 
%The contour plot of out-of-plane magnetic field with T=10eV is shown in the figure(4) and can be seen that the instability rapidly grows with the inclusion of the temperature.

\begin{figure}
\centering
\includegraphics[scale=0.65]{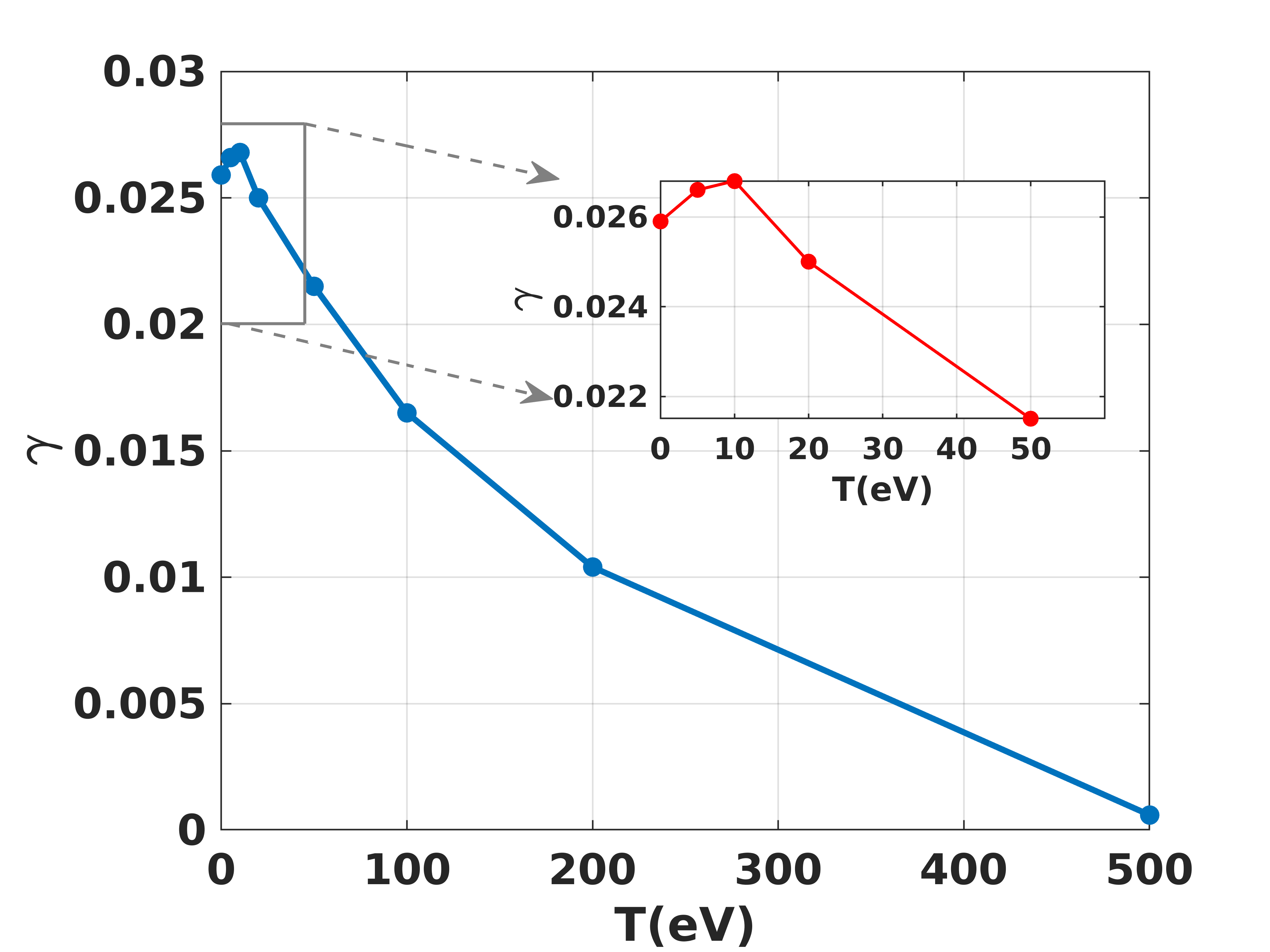}% Here is how to import EPS art
\caption{\label{fig:pt_GR vs T}Variation in growth rate as a function of temperature (T) in eV, calculated from the perturbed energy evolution for the pure tearing mode case ($C_0=0$). The inset presents a closer look at the growth rates for temperatures of 0, 5, 10, 20, and 50 eV.}    
\end{figure}

\begin{figure}
    \centering
    \includegraphics[scale=0.65]{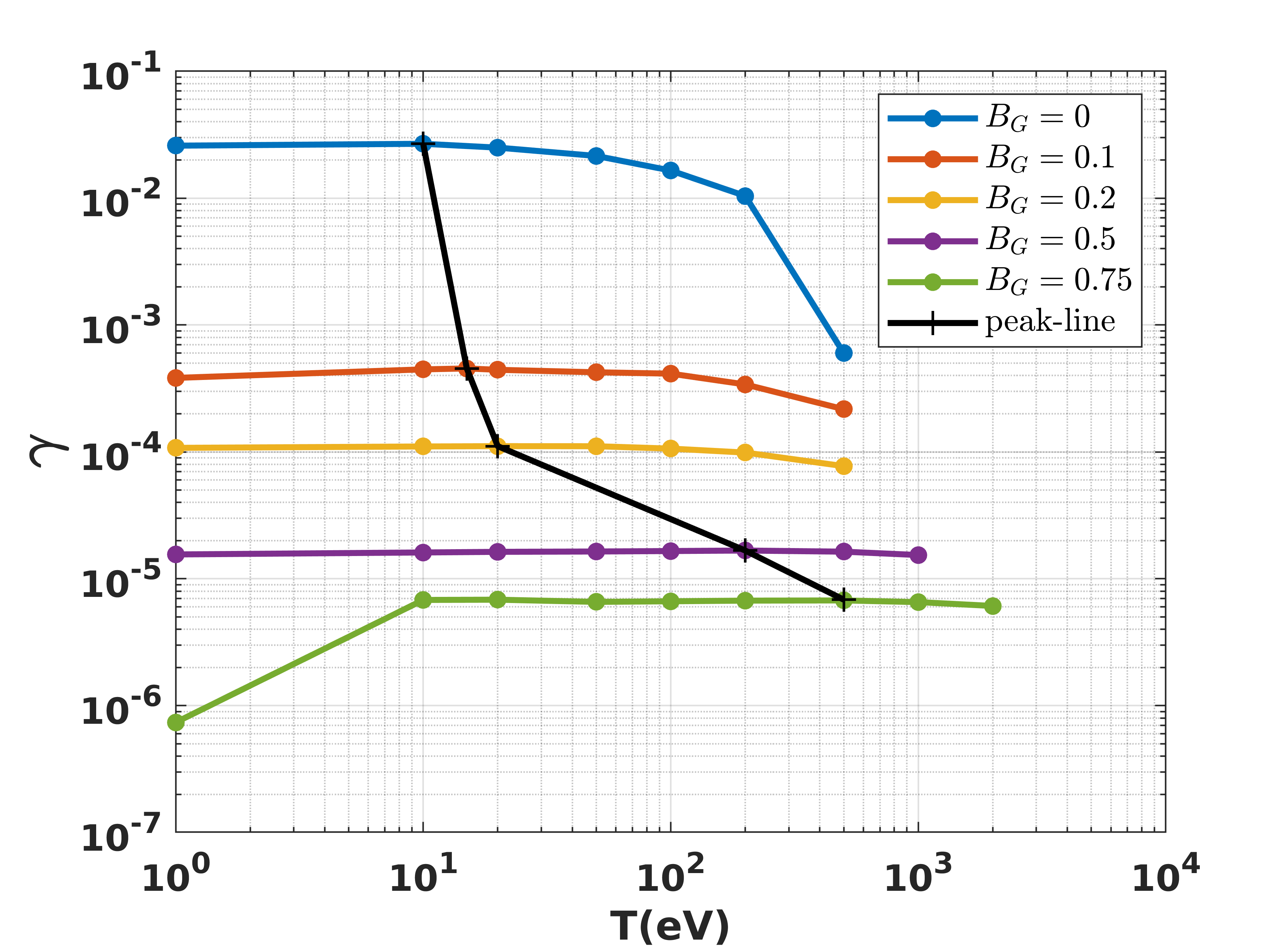}
    \caption{Growth rate of perturbed energy evolution for the pure tearing mode case ($C_0=0$) with different guide field values and varying temperatures. The black line denotes the peak value of each growth rate curve for the different guide field values.}
    \label{fig:guide field}
\end{figure}

\subsection{$C_0 = 0.06$: Only Surface Preserving Mode Case }
We now add \( C_0 \), a uniform magnetic field aligned with the equilibrium magnetic field. Here, \( B_0 \) represents the equilibrium magnetic field, and \( B_0'' = d^2B_0/dx^2 \) is its second derivative.The quantity \( B_0 - B_0''\) must change sign for instabilities to exist, a necessary condition highlighting the role of the current gradient. The condition \( B_0 B_0'' \leq 0 \) is a well-known criterion for the tearing instability, as discussed in the literature \cite{bulanov1992}. Conversely, the condition \( B_0 B_0'' > 0 \) corresponds to the existence of the surface-preserving mode, as detailed by Gaur and Kaw \cite{gaur2016}. 

In our setup, we choose \( C_0 = 0.06 \), which ensures that the condition \( B_0 B_0'' \leq 0 \) for tearing instability is violated. At the same time, the condition \( B_0 B_0'' > 0 \) is satisfied, favoring the surface-preserving mode. As a result, only the surface-preserving mode is present in this configuration.

Simulation run with $T = 0~eV$, show no formation of quadrupoles or islands, which is unlike the tearing mode scenario. Magnetic field lines in this case do not bend and instead form a channel-like structure, as depicted in Fig.~\ref{fig:sp_streamlines_0eV}. In order to understand the behaviour of growth rate of surface preserving mode we conduct simulation runs for the following temperature values: $( T = 5, 10, 20, 50, 100, 200, 500)$ electron volts. In Fig.~\ref{fig:sp_GR vs T}, we plot the growth rate vs temperature. In this case, the growth rate goes on to increase, unlike the tearing mode case where the growth rate first increases and then start decreasing after a certain temperature. This can be explained as follows. Surface preserving mode, also termed as stationary nontearing inertial scale mode, is driven by the electron flow perturbations along the direction of the current gradient\cite{lukin2009}. An increase in temperature increases the thermal velocity of electrons, which enhances their mobility and indirectly amplifies the electron flow perturbations, making the instability more favorable.

\begin{figure}
\centering
\includegraphics[scale=0.90]{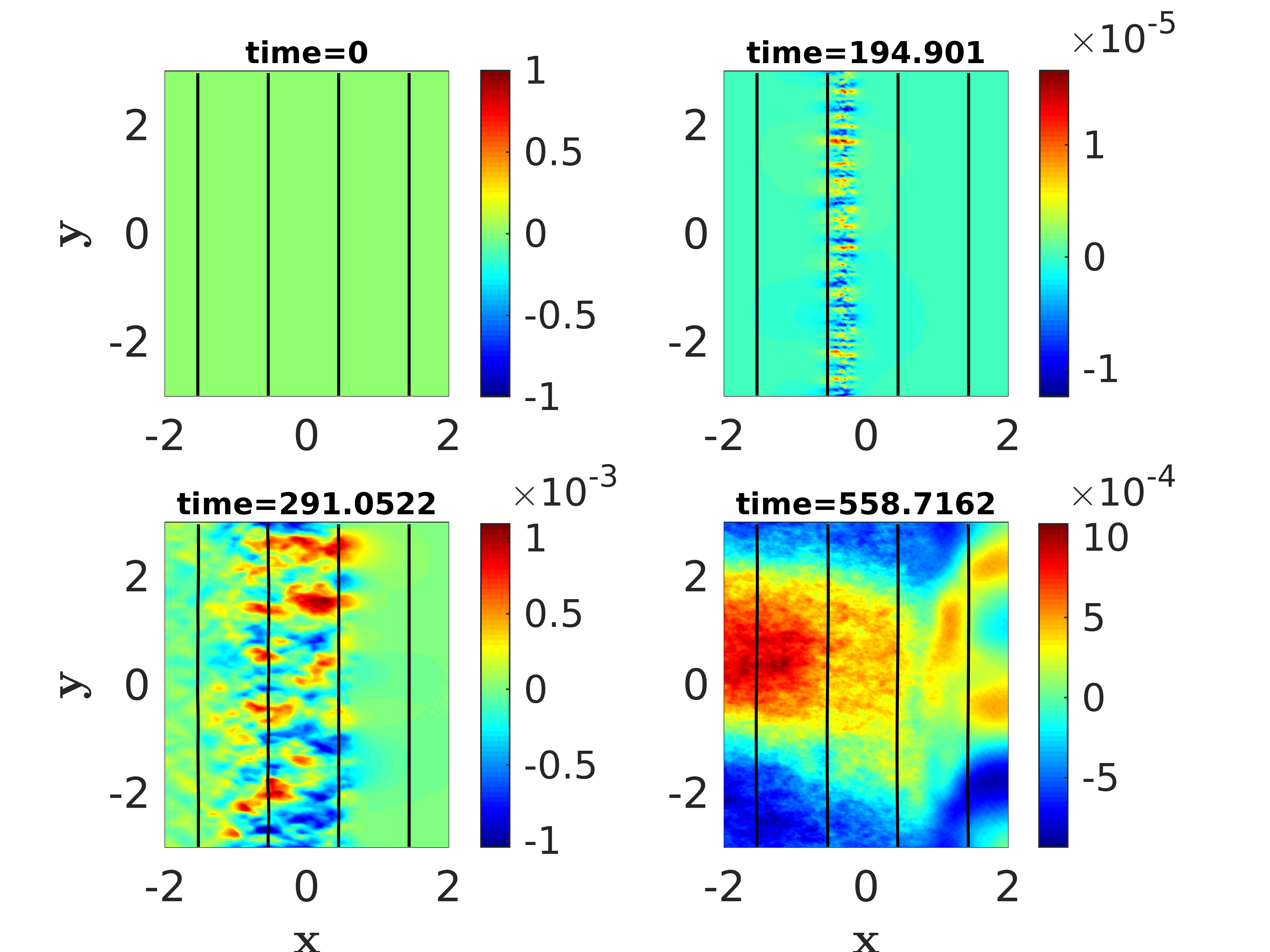}% Here is how to import EPS art
\caption{\label{fig:sp_streamlines_0eV} Shaded isocontours of out-of-plane magnetic \(B_z\) field with superimposed streamlines (solid lines) over various times for the only surface preserving mode case ($C_{0}=0.06$) and T = 0 eV case. The color bar indicates the normalized amplitude of \(B_z\).}
\end{figure}

\begin{figure}
\centering
\includegraphics[scale=0.65]{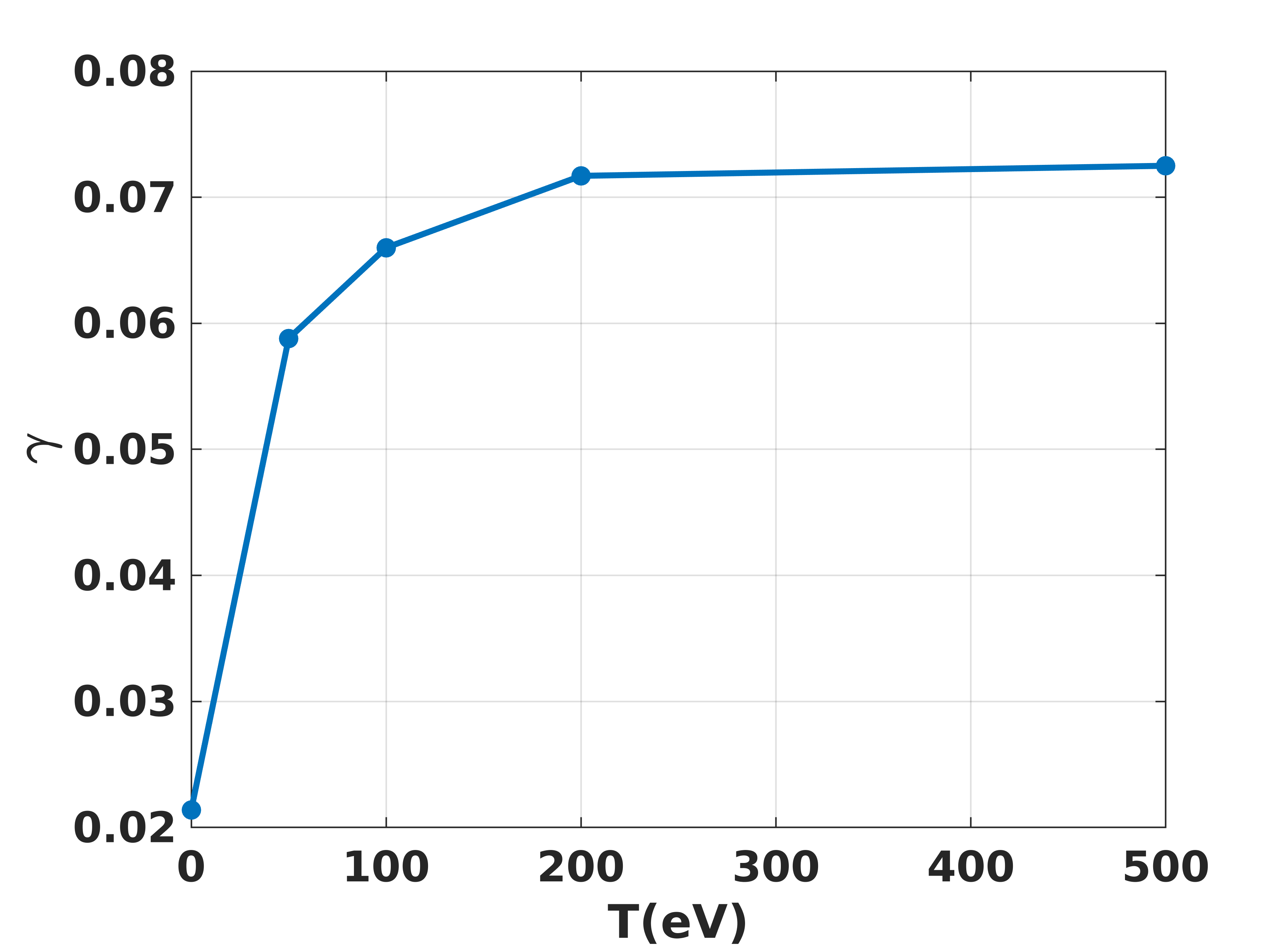}% Here is how to import EPS art
\caption{\label{fig:sp_GR vs T} Variation in growth rate as a function of temperature (T) in eV, calculated from the perturbed energy evolution for the only surface preserving mode case ($C_0=0.06$).}
\end{figure}

\subsection{$C_0 = 0.025$: Mixed Modes Case}
In this last case, we choose $C_0 = 0.025$. This value of $C_0$ permits both tearing and surface preserving modes to exist simultaneously. Due to the presence of tearing mode, the initially straight magnetic field lines evolve and then form an island. However, unlike the only tearing mode case (for $C_0 = 0$), the island formed in this case is asymmetric (see Fig.~\ref{fig:mm_streamlines_0eV}). The asymmetry is also observed in the evolution of out-of-plane magnetic field, the quadrupole formed is asymmetric.

Another interesting observation we made is that the island does not remain at its original location; it moves towards the left, the side with the weaker magnetic field. In Fig.~\ref{fig:mm_jz}, we see that initially the island forms at the null-line, denoted by a vertical line (at x = - 0.2661 ), and it is asymmetric. As time progresses, the island shifts leftward. As seen in last subplot, the island has moved away from the vertical null-line. In contrast, in the case of only the tearing mode, a symmetric island forms at the null-line (at x = 0.0 in this case) and remains there, as shown in various subplots of Fig.~\ref{fig:pt_jz}. Shaded isocontours of equilibrium current show the formation of mushroom-like patterns. Two jet-like electron flows emerge from the neighboring $X$-points, which are essentially the outflows from two adjacent reconnection sites. These flows collide at the center of the $O$-point and then move perpendicularly to impact the closed magnetic surfaces on opposite sides, creating the mushroom-like structures. In the mixed modes case, the two electron flows encounter different strengths of the magnetic field on either side of the null-point. The stronger magnetic field experiences less bending due to higher magnetic tension. The collision of electron flows with magnetic surfaces of unequal strengths results in a net momentum toward the left. This net momentum causes the island and other structures to move in the negative x-direction. The unequal magnetic field strength on the two sides of the null-line is also the reason behind the formation of asymmetric structures. Some of these features have been reported in studies on asymmetric magnetic reconnection, which has recently garnered significant interest.

To understand the role of temperature, we conducted simulation runs for the following temperature values: T = 5, 10, 20, 50, 100, 200, 500 electron volts. Figure~\ref{fig:mm_GR vs T} depicts the change in growth rate with temperature for this case. Similar to the case with only the tearing mode, the growth rate first increases with temperature and then starts decreasing. However, the peak of the curve has shifted to a lower temperature. Specifically, the peak growth rate occurs around 5 eV, compared to 10 eV in the only tearing mode case. The reason for this leftward shift of the peak is not certain at present. It could be due to asymmetry in the magnetic field on two sides of the null-line, the co-existence of the surface-preserving mode, and the effect of temperature on them. A more detailed study is underway to understand this effect.

\begin{figure}
\centering
\includegraphics[scale=0.90]{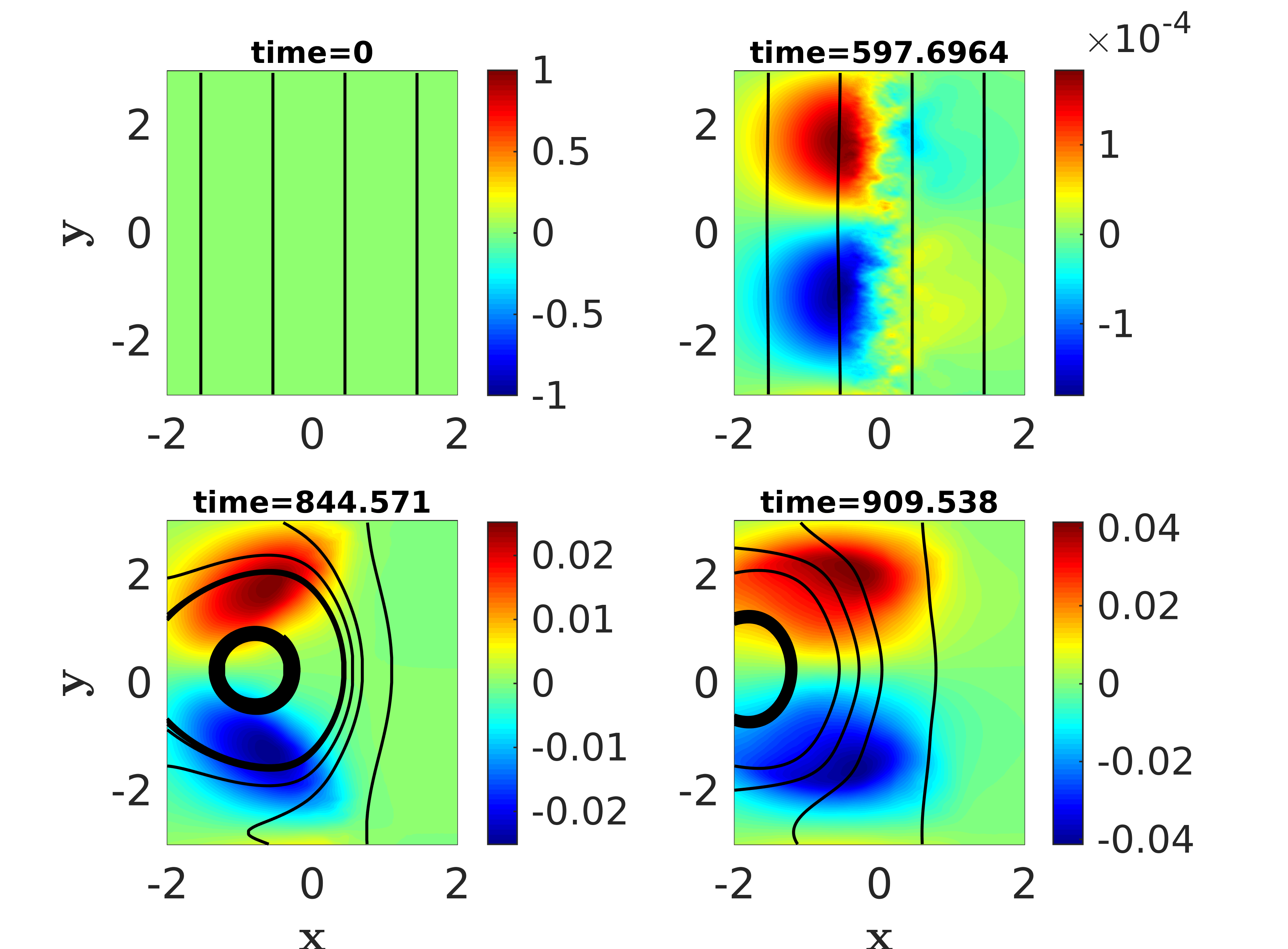}% Here is how to import EPS art
\caption{\label{fig:mm_streamlines_0eV} Shaded isocontours of out-of-plane magnetic field \(B_z\) with superimposed streamlines (solid lines) over various times for mixed-modes case ($C_{0}=0.025$) and T = 0 eV case. The color bar indicates the normalized amplitude of \(B_z\).}
\end{figure}

\begin{figure}
\centering
\includegraphics[scale=0.90]{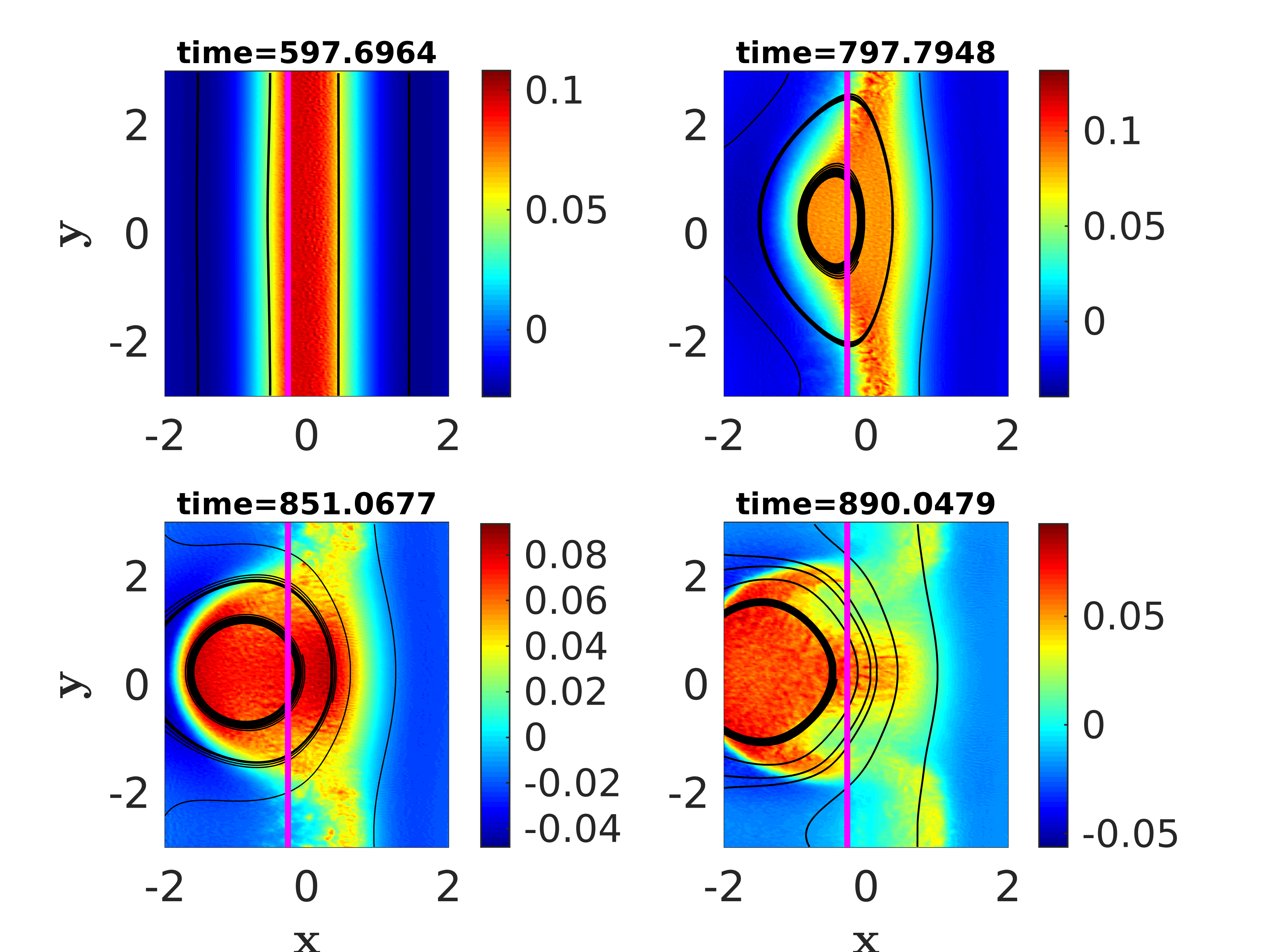}% Here is how to import EPS art
\caption{\label{fig:mm_jz} Shaded isocontours of the current $J_z$ with superimposed streamlines (black lines) over various times for mixed-modes case ($C_{0}=0.025$) and T = 0 eV case. The color bar indicates the normalized amplitude of \(J_z\).}    
\end{figure}

\begin{figure}
\centering
\includegraphics[scale=0.90]{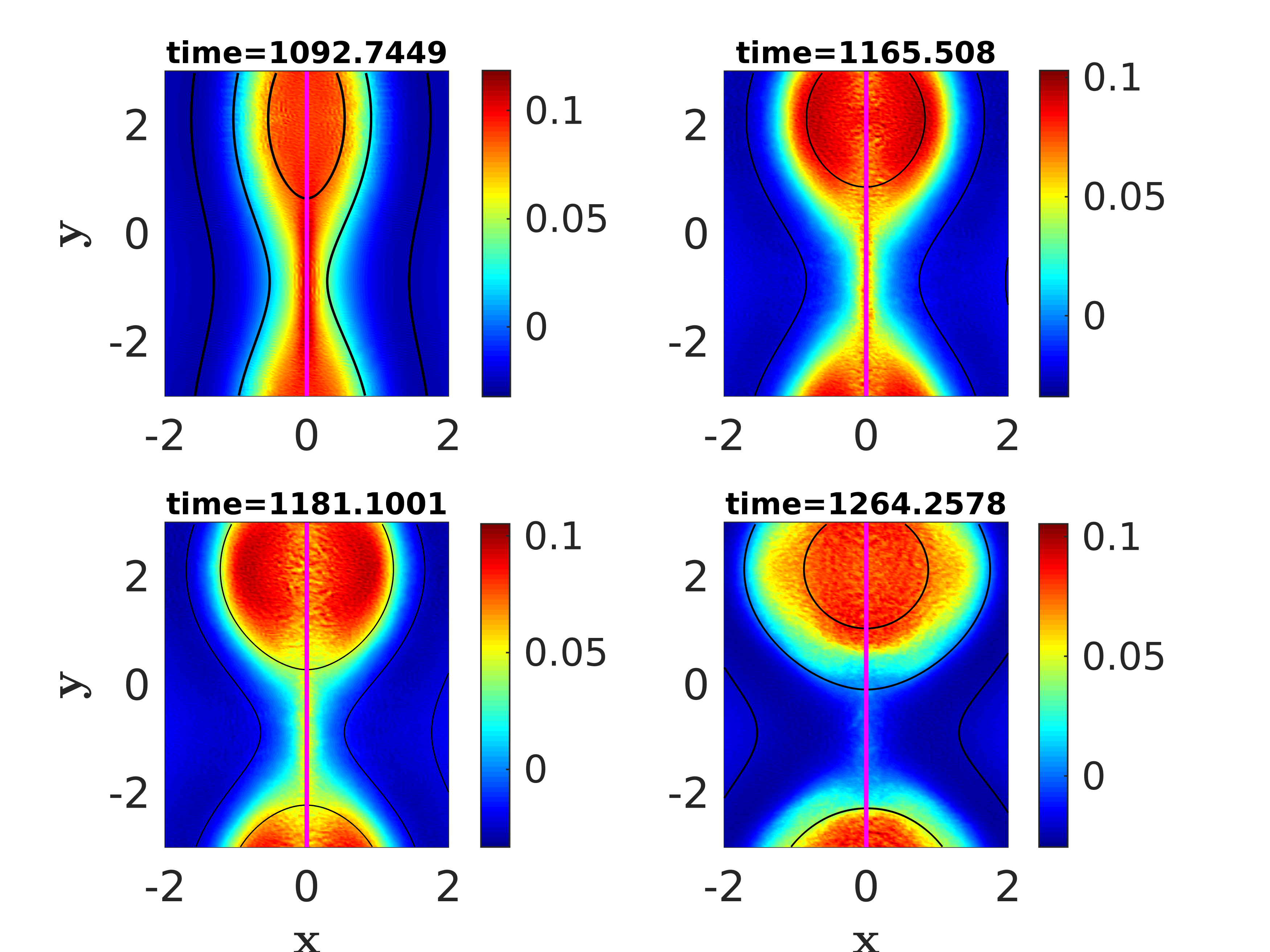}% Here is how to import EPS art
\caption{\label{fig:pt_jz} Shaded isocontours of the current $J_z$ with superimposed streamlines (black lines) over various times for pure tearing mode case ($C_{0}=0$) and T = 0 eV case. The color bar indicates the normalized amplitude of \(J_z\).}  
\end{figure}

\begin{figure}
\centering
\includegraphics[scale=0.65]{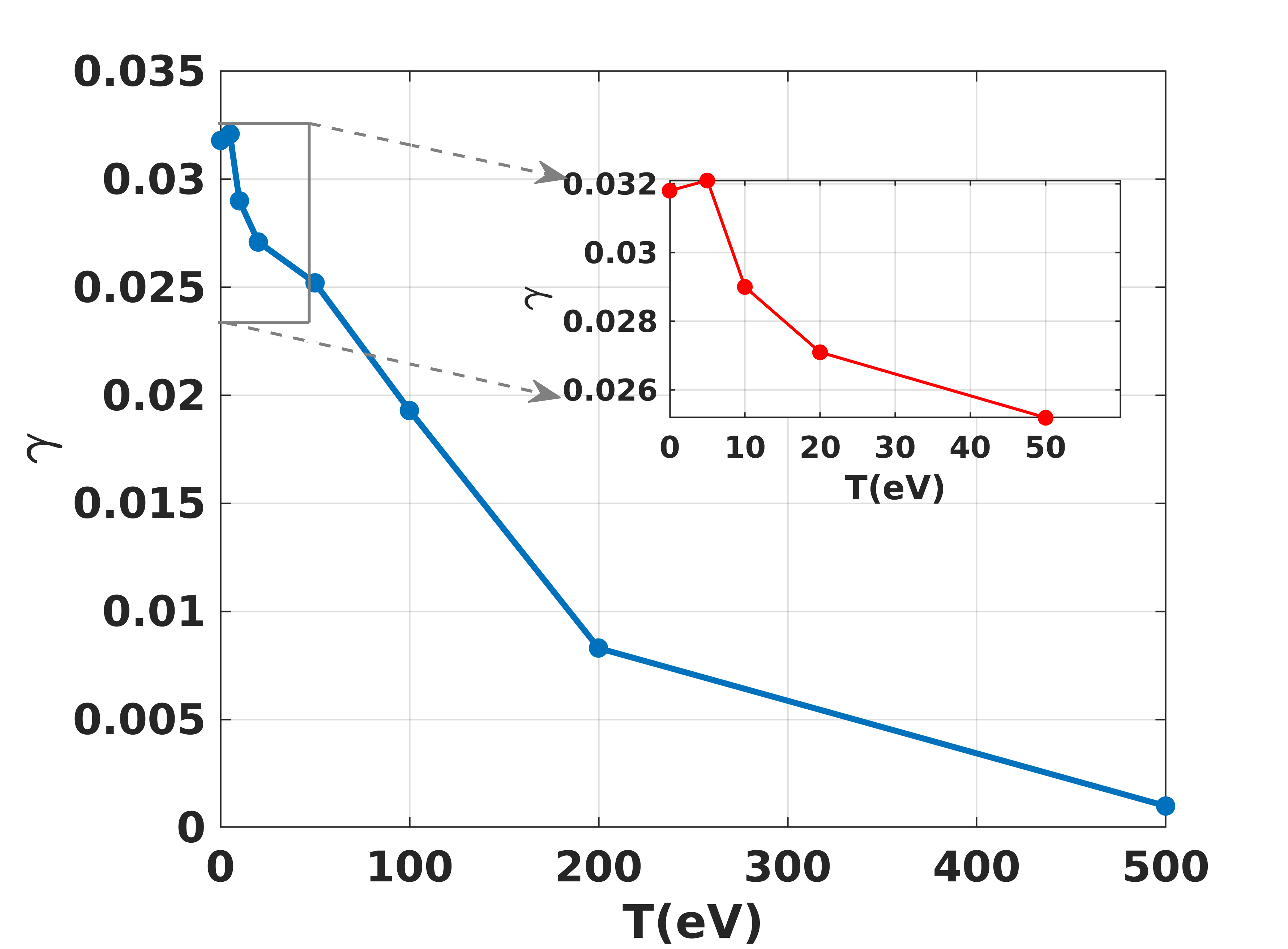}% Here is how to import EPS art
\caption{\label{fig:mm_GR vs T} Variation in growth rate as a function of temperature (T) in eV, calculated from the perturbed energy evolution for the mixed-modes case ($C_0=0.025$). The inset presents a closer look at the growth rates for temperatures of 0, 5, 10, 20, and 50 eV.}    
\end{figure}

\subsection*{Comparison with the Existing Studies:}

Resistive MHD model can not explain the fast reconnection. Collisionless reconnection is one of the most promising models to explain the same. A vast literature is available on collisionless reconnection where hall term and inertia term become important. But most of these studies include the ion motion as well. Electron scale tearing mode is considered to be a prominent mechanism behind the fast reconnection. Studies that focus on the motion of only electrons by assuming the ions to be at rest, like our studies presented in this paper. We compare our results with the previous studies those have considered electrons motion only.

Electron scale tearing instability has been studied using Electron Magnetohydrodynamics (EMHD) model and Particle-In-Cell (PIC) simulations. Cai and Li \cite{cai2009}, studied the pressure gradient effects on tearing mode in the presence of guide magnetic field by including the electron pressure tensor, using EMHD model. Our numerical simulations are consistent with their analytical studies which show that when the electron thermal Larmor radius becomes sufficiently large, it can suppress the tearing mode instability. In one of their earlier works, Cai and Li \cite{cai2008}, had included the compressible effects. However, our PIC simulations do not include the compressible effects and focus on the kinetic effects and specifically examine how temperature variations influence the tearing mode. Studies by Cai et al., are limited to linear regime only, while our simulations explore the nonlinear regime as well.

Jain \& Büchner \cite{jain2015} conducted 3D linear studies focusing on tearing and non-tearing modes, examining the dependence of growth rate on the thickness of the electron current sheet, the strength of the guide field, and the wavenumber of the modes. Later, Jain et al. \cite{jain2017} performed 3D simulations to study these effects in the nonlinear regime. The key difference in our work is the inclusion of temperature effects, with no perturbations assumed along the equilibrium velocity current.

Guo et al. \cite{guo2020} analyzed the density and pressure gradient effects on the tearing mode in the context of 2D EMHD to study the electron diamagnetic drift and Biermann battery effects. In their later studies, Guo et al. \cite{guo2021} explores the tearing mode instability within electron-skin-depth-scale current sheets using EMHD model. The study conducts linear simulations of the tearing mode in both Harris sheet and force-free current sheet configurations, revealing that resistive diffusion is negligible at this scale. Additionally, the study examines the impact of a strong magnetic guide field and shear flow, finding that shear flow can suppress the tearing mode in a force-free equilibrium state. Our research, on the other hand, does not consider density and pressure gradients or shear flow, but rather focuses on the interplay between electron temperature and the guide field.

The contour plots for magnetic field, current and other fields observed in our PIC simulations exhibit the same behavior as observed by Sarto et al. \cite{sarto2005} during the fully nonlinear phase of the EMHD reconnection. We have also observed the secondary electron Kelvin – Helmholtz instability of the current jets originating from two nearby X points colliding at the O point. However, in the case of mixed mode, we observe that the collision of the two jets results in the net momentum and make the island and other structure move from the position they initially form. No such phenomena was observed in the case of only surface-preserving mode case.

Betar et al. \cite{betar2024} conducted a numerical study on the linear dynamics of tearing modes within a slab incompressible EMHD framework, accounting for the influence of more than one non-ideal parameters such as electron inertia, resistivity and electron viscosity. Their findings emphasize the need for future research to incorporate kinetic effects and investigate EMHD tearing mode regimes, such as “electron-only” reconnection. 

Zhao \cite{zhao2017} incorporated the temperature in the warm EMHD model and derived the dispersion relation and investigated the electromagnetic properties of obliquely propagating whistler waves. As a future work, the warm EMHD model discussed by Zhao can be used to study linear growth rate and eigen structure of tearing and surface preserving modes change at low to moderate temperature (at high temperature, the EMHD model would break). The work on the same is underway.

\section{Summary and Future Scope}
In this study, we conducted two-dimensional PIC simulations using the OSIRIS framework to investigate the kinetic effects on tearing and surface-preserving instabilities in electron current layers, which are relevant to various space, astrophysical, and laboratory plasma environments. We started by simulating the conventional tearing instability to validate our simulation setup. After validation, we explored the kinetic effects on the tearing mode by varying the electron temperature. Our results show that the growth rate of the tearing instability initially increased with temperature, peaking around 10 eV, before declining due to the increased electron Larmor radius and consequent diffusion of magnetic fields. To support our argument, we introduced a uniform guide field, which exhibited a stabilizing effect on the tearing mode, resulting in a shift of the peak growth rate to higher temperatures.

To study the surface-preserving instability, we added a uniform magnetic field \( C_0 = 0.06 \) along the equilibrium magnetic field direction, which inhibits the tearing mode in the system. To analyze the effect of temperature on the growth rate of this mode, we ran simulations at various temperatures. The growth rate versus temperature plot shows a consistent rise, which can be attributed to the increased electron flow perturbations at higher temperatures. This rise favors the instability, as it is driven by perturbations in the flow. 

Finally, we also investigated the case when both the tearing and surface-preserving modes exist simultaneously in the system. Simulation runs in this scenario show that the structures (island and quadrupole) formed are asymmetric. Moreover, these structures do not remain stationary but move towards the weaker magnetic field side around the null-line. This behavior aligns with findings from studies on asymmetric magnetic reconnection. The growth rate's temperature dependence follows a pattern similar to the pure tearing mode case, initially increasing with temperature but peaking at lower temperature value. The causes of this shift are currently being studied, considering magnetic field asymmetries and the interplay between tearing and surface-preserving modes under varying temperatures.
\subsection*{{\it Future Directions:}}
\begin{itemize}
    \item In our studies, we have chosen the tanh - sech profile for the equilibrium magnetic field. While this allows us to use periodic boundary conditions, the profile naturally ceases at the boundaries. Conducting studies on larger systems (using profiles with a larger extent) with appropriate boundary conditions would provide a natural comparison.
    \item In our studies, we have chosen \(\epsilon = 1\), which is normalized by the electron skin depth. A parametric study by choosing \(\epsilon > 1\) and \(\epsilon < 1\) needs to be carried out to understand the behavior of the two modes in these different regimes, along with the role of temperature on their behavior.
    \item To focus on electron dynamics, we approximated the ions as stationary, thereby neglecting ion motion. While this approach simplifies the analysis, future work will incorporate mobile ions to explore their dynamic effects, such as the electron-cyclotron drift instability and the interplay with additional magnetic field components.
    \item In mixed mode cases, the magnetic field profile is asymmetric around the null line. To distinguish the effects of this asymmetry on the tearing mode and its interaction with the surface-preserving mode, a dedicated study using appropriate profiles is needed. 
    \item Linear stability analysis using a warm EMHD (Electromagnetic Magnetohydrodynamics) model is currently underway and is essential for understanding the eigenmode structure under varying temperature conditions. This analysis will provide insights into how temperature influences the structure and stability of the modes. 
\end{itemize}

\subsection{Acknowledgments}
SM gratefully acknowledges the financial support provided by the Govt. of Gujarat under SHODH scheme. SM and GG would like to thank Professor Abhijit Sen for useful discussions. GG would like to thank Jugal Chowdhury for providing the comments on the manuscript.

\section*{Appendix: Role of Current Gradient}

The current gradient serves as the free energy source driving tearing and surface-preserving instabilities. 

In the absence of ion motion (i.e., \( v_i = 0 \)), the current density is: 

\[
\mathbf{J} = - n_e e \mathbf{v}_e,
\]

If \( n_e \) is uniform, as in our case, the gradient in the equilibrium current becomes:

\[
J_e' = \frac{dJ_e}{dx} = - n_e e v_e',
\]

where \( v_e' \) is the spatial gradient in the equilibrium electron velocity. 

For a one-dimensional current configuration, the current gradient \( J_e'(x) \) is inversely proportional to the spatial scale of current variation, as expressed by:

\[
J_e' = \frac{\Delta J_0}{\epsilon},
\]

where \( \epsilon \) represents the spatial scale. Thin current configurations (such as current sheets) represent sharper current gradients, which are critical in driving instabilities. Maxwell’s relation gives:

\[
J_e' = B_0'' \left( = \frac{d^2 B_0}{dx^2} \right) = - n_e e v_e',
\]

The one-dimensional equilibrium magnetic field, corresponding to a thin current sheet, aligns in opposite directions across the current sheet. Non-ideal effects such as resistivity, electron inertia, and turbulence cause the magnetic field to decay, inducing plasma motion to fill the space created by the decaying fields.

The quantity \( B_0 - B_0'' \) (or equivalently \( B_0 + v_e' \)) must change sign for instabilities to exist, a necessary condition highlighting the role of the current gradient \cite{gaur2016}.

\section*{References}
%\begin{thebibliography}{<num>}
\bibliographystyle{iopart-num} % Replace 'yourstyle' with the name of your .bst file (without the .bst extension)
\bibliography{iopart-num} % Produces the bibliography via BibTeX.
%\end{thebibliography}

\end{document}